\documentclass[prd,superscriptaddress,nofootinbib,showkeys, preprint]{revtex4}
\usepackage{graphicx}
\usepackage[usenames,dvipsnames]{color}
\usepackage{amsmath,amssymb}
\usepackage{mathtools}
\usepackage[colorlinks]{hyperref}
\usepackage{float}

\newcommand{\be}{\begin{eqnarray}}                                             
\newcommand{\ee}{\end{eqnarray}}
\newcommand{\nn}{\nonumber}

\begin{document}

\title{ Equivalence of two solutions of physical massive gravity}

\author{Haresh Raval}
	\email{haresh@phy.iitb.ac.in}
	\author{Krishnanand Kr.Mishra}
		\email{ mishra.krishna93@gmail.com}
		\author{Bhabani Prasad Mandal}
	 \email{ bhabani.mandal@gmail.com}
	
	\affiliation{Department of Physics, Institute of Science, Banaras Hindu University,Varanasi-221005, India}

\begin{abstract}
We study some aspects of linearized gravity as gauge theory, with massive deformation. Recently it has been shown that there are two distinct solutions, which represent physical massive gravity. The purpose of the present work is to show that these two seemingly discrete solutions are equivalent at the level of generating functional. The significance of this simple work lies in the fact that one solution represent physical massive gravity at Fierz-Pauli (FP) point and other outside the FP point. 
\end{abstract} 



\maketitle 

\section{Introduction}
Despite the fact that general relativity (GR) remains a beautiful and accurate theory, there has been recent interest in the large distance modification of GR. One of the motivations comes from the observation of supernova data~\cite{1,2} which shows that the universe has  an accelerating rate of
its expansion. If GR is correct, there must exist some dark energy density,  $\rho\sim 10^{-29} g/cm^3$ to explain the accelerating expansion.  This in turn implies that
 there is a constant term, $\Lambda$, in the Einstein-Hilbert (EH) action,
which would give $\rho\sim M^2_P \Lambda$.  To give the  vacuum energy which can give the correct rate of expansion at least up to the order,  this constant has to be of the small value whereas the quantum field theory suggest a value
much larger \cite{3}.  Therefore, this fact prompts one to think that 
 may be gravity itself gets modified at large distances.  One of the modifications is to give graviton the mass.  The mass of the graviton suppresses effect of large cosmological constant to produce the rate of expansion which coincides with the observations~\cite{4}. This mechanism is known as degravitation \cite{5,6,7,8}.  The massive gravity however breaks diffeomorphism invariance of the GR.  This is consistent with the analysis coming from the holographic approach to the study of thermo-electric transport in condensed matter \cite{9,10,11}.  There, the momentum dissipation results from the translational invariance breaking due to impurities. The theory of massive gravity represents a way to implement momentum relaxation holographically \cite{12}.  Giving graviton the mass is technically non-trivial phenomenon, it changes the degrees of freedom (dof) of the graviton from $2$ to $5$.  Hence,  physical massive gravity must have 5 dof.  There are other modifications to GR that one can cook up such as replacing EH action with $f(R)$ gravity,  a general function of the Ricci scalar \cite{13,14}, which can lead to self-accelerating solutions without the need of the cosmological constant \cite{15,16}.  Inconsistency with the cosmological constant is not however the only motivation to modify the gravity at large distances. There are other good provocations~\cite{4,9,10,11,12,17,18,19} too for such study.
 
 In this paper,  we focus on perturbative modifications,  given by symmetric tensor $h_{\mu \nu}(x)$ around the background Minkowski metric $\eta_{\mu \nu}(x)$, in terms of which the linearized Einstein-Hilbert (EH) action is built. The theory of linear perturbations in EH gravity  is treated as the gauge theory of the rank 2 tensor field.  The characteristics of the graviton in this theory are described by $h_{\mu \nu}(x)$.  We 
consider massive gravity with the most general Lorentz invariant mass term.  There are many mathematical solutions for such a model of massive gravity which exhibit absence of tachyonic poles in the propagator. However, only two of them,  one on the so called FP point and the other outside FP point correspond to physical theory of massive gravity.  The aim of this paper is to show the equivalence of two physical massive gravity solutions given in the Refs. \cite{20,21} which require well defined propagators.  We need to fix the gauge properly to have well defined propagators for graviton. The gauge fixed action is then made BRST invariant.

We shall take recourse to a well known  technique  of finite field dependent transformations(FFBRST) to establish the equivalence of two solutions.  FFBRST was  introduced  for the first time in Ref. \cite{22}. Over the period it has found various applications in field theory~\cite{23,24,25,26,27,28,29, 30,31,32,33}. In a recent interesting work, the FFBRST technique itself has been extended to include connection of the theory with  two different gauge fixings with a theory having only Lorenz gauge fixing~\cite{34}. FFBRST transformation is a generalization of usual BRST transformation, with the infinitesimal  global anti commuting parameter being replaced by a finite parameter dependent on fields. Such a finite generalization of BRST transformation preserves nilpotency and retains the original BRST symmetry of gauge theory. Due to finiteness and field dependence of the parameter,  the path integral measure does not remain invariant  under FFBRST which results in Jacobian contribution in path integral.  The Jacobian depends on choice of field dependent parameter which under certain conditions can be represented as the local functional of fields and can be a part of the action.
Using such appropriately constructed FFBRST parameter, one  can relate the generating functionals of two effective theories \cite{22}. In this present work, we connect generating functionals which correspond to two physical massive gravity solutions, one at FP point and the other outside FP point to show their equivalence.

 The paper is organized as follows: in the next section we review the massive gauge fixed linearized gravity. In section III, the possible candidates for the physical massive gravity are listed.  In section IV, the FFBRST technique is employed as the tool to demonstrate the equivalence of two solutions. Last section is kept for conclusions.
\section{ review of linearized massive gravity}
Here we study the linearized massive gravity with massive deformation in $D$ space-time dimensions. The most general action of the linearized gravity in $D$ dimensions is given by 
\begin{equation}\label{1}
S_{inv}=\int d^Dx
\left(
\frac{1}{2}h\partial^2h - h_{\mu\nu}\partial^\mu\partial^\nu h - \frac{1}{2}h^{\mu\nu}\partial^2h_{\mu\nu} + h^{\mu\nu}\partial_\nu\partial^\rho h_{\mu\rho}
\right),
\end{equation}
where  $h_{\mu\nu}(x)$  is a  symmetric tensor of rank-2,  $h(x)$ is trace of $ h_{\mu\nu}(x):h(x)=h^{\lambda}_{\lambda}=\eta^{\mu\nu} h_{\mu\nu}$ and  $\eta_{\mu\nu}=diag(-1,1....1)$. The action \eqref{1} is invariant under the following gauge transformation
\begin{equation}
\delta h_{\mu\nu}=\partial_\mu\theta_\nu + \partial_\nu\theta_\mu.
\end{equation}
{\bf
{ Gauge fixed Model :}
}
Since we require to have well defined propagator for graviton in this theory, we now choose the following gauge fixing
\begin{equation}
\partial^\nu h_{\mu\nu} +\kappa_1\partial_\mu h=0.
\end{equation}
It is at once clear that it is the analog of $\partial^\mu A_\mu = f(x)$ in the vector gauge theory. By the usual Faddeev-Popov procedure, this condition amounts to the following gauge fixing and ghost actions
\begin{eqnarray}\label{2}
S_{gf}+S_{ghost}=\int d^{D}{x}[b^\mu (\partial^\nu h_{\mu\nu}+\kappa _1\partial_\mu h)+\frac{\kappa}{2} b^\mu b_\mu + \partial^\nu \overline{\xi}^\mu ((1 + 2\kappa_1)\partial_\mu \xi_\nu + \partial_\nu \xi_\mu]
\end{eqnarray}
where $\xi$, $\overline{\xi}$ are ghost, anti-ghost respectively and $b_\mu$ is a Nakanishi-Lautrup Lagrange multiplier.
From the Eq.~\eqref{2},  we see that there are two gauge parameters,  $\kappa$  and $\kappa_1$.  Therefore, effective action of theory is given by
\be\label{3}
S_{eff}= S_{inv}+S_{gf}+S_{ghost}
\ee
 Now the gauge fixed action has well defined propagators and is invariant under the following  BRST  transformation,
 \begin{eqnarray}\label{4}
 	sh_{\mu\nu}&=&(\partial _\mu \xi_\nu + \partial_\nu\xi_\mu )\delta\omega\nn\\
 	s\xi_\mu &=&0 \\
 	s\bar \xi_\mu & = &b_\mu \delta\omega \nn\\
 	s b_\mu & =&0. \nn
 \end{eqnarray}
  As we are interested in massive gravity, we now introduce the mass term in the theory given in Eq.~\eqref{3} as follows, 
\begin{equation}
S_m=\int d^Dx \frac{1}{2} \left (
m_1^2h^{\mu\nu}h_{\mu\nu} + m_2^2 h^2
\right)
\end{equation}
where $m_1^2$  and $m_2^2$ are  mass parameters for $h_{\mu\nu}(x)$ and $h(x)$, respectively. This is the most general Lorentz invariant mass term in D dimensions.
It is easy to see that the mass term breaks the BRST invariance under transformations~\eqref{4}. The masses $m_1^2$ and $m_2^2$ are not free parameters and are restricted to have physical theory of massive gravity.

\section{Massive gravity solutions}
In order to have a solution for physical massive gravity, there are two requirements for it to simultaneously satisfy: \\
(1) Absence of unphysical tachyonic poles in the propagator.\\
(2) Degrees of freedom (dof) of graviton must be five in four dimensions.\\
 The former requirement gives us eight possible candidates for a given set of parameters as we discuss now. The
massive theory is given by the action
\be\label{S}
 S'= S_{eff}+S_{m}.
\ee
 We see that it depends on four parameters: $m^2_1, \ m_2^2, \ \kappa$ and $\kappa_1$ ~\cite{20,21}. 
There are only four possible combinations  for two gauge fixing parameters given as below ~\cite{20,21}:
	\begin{itemize}
	\item[a.]  $\kappa\neq 0, \ \kappa_1=0$
	\item [b.] $\kappa\neq 0, \ \kappa_1=-1$
	\item[c.]  $\kappa = 0, \ \kappa_1=0$
	\item [d.] $\kappa = 0, \ \kappa_1=-1$
	\end{itemize}

\noindent Requesting the absence of unphysical poles in the propagators, it is found that there are eight different  
constraints on  masses $m_1^2$ and $m_2^2$ in total depending upon the combination of two gauge parameters \cite{20,21}. Each constraint corresponds to a solution. These solutions are 
\begin{description}
\item[solution 1] 
\begin{equation}\label{3.4}
m_1^2> 0 \ ;\ m_1^2+Dm_2^2= 0\ ;\ \kappa<0\ ;\ \kappa_1=0.
\end{equation}
\item[solution 2] 
\begin{equation}
m_1^2>0\ ;\ 
m_1^2+Dm_2^2<0\ ;\ 
\kappa\leq\frac{D-1}{2-D}\ ;\ \kappa_1=-1
\label{3.5}\end{equation}
\item[solution 3] 
\begin{equation}
m_1^2>0\ ;\ 
m_1^2+Dm_2^2>0\ ;\ 
\frac{D-1}{2-D}\leq\kappa<0\ ;\ \kappa_1=-1
\label{3.6}\end{equation}
\item[solution 4] 
\begin{equation}
m_1^2=0\ ;\
m_2^2<0\ ;\ \kappa\leq\frac{D-1}{2-D}\ ;\ \kappa_1=-1
\end{equation}
\item[solution 5] 
\begin{equation}
m_1^2=0\ ;\
m_2^2>0\ ;\ 
\frac{D-1}{2-D}\leq\kappa<0\ ;\ \kappa_1=-1
\label{3.8}\end{equation}
\item[solution 6] 
\begin{equation}
m_1^2=0\ ;\
m_2^2<0\ ;\ \kappa>0\ ;\ \kappa_1=-1
\label{3.9}\end{equation}

\item[solution 7] 
\begin{equation}
m_1^2 \geq 0 \ ;\ m_1^2+Dm_2^2 \leq 0\ ;\ \kappa=0\ ;\ \kappa_1=0
\label{3.10}\end{equation}
\item[solution 8] 
\begin{equation}\label{3.11}
m_1^2\geq 0 \ ;\ m_1^2+Dm_2^2\neq 0\ ;\ \ ;\ \kappa=0\ ;\ \kappa_1=-1 
\end{equation}
\end{description}
$D$ is the spacetime dimension of theory.
From solution 2 to solution 6 are five different cases of combination $b$.
 We note that ranges of masses in five amongst eight solutions namely 1, 3, 4, 5 and 6,  exclude the FP point 
\be
m_1^2 + m_2^2=0.
\ee
Significance of this point is at once clear in the model of massive gravity without any gauge fixing. The model will correspond to a physical theory only at this FP point i.e.,  graviton in that model has $5$ dof only if FP condition is met.
The question now of course is, amongst the eight solutions listed above in the gauge fixed model, which represent $also$ a physical theory of massive gravity,  $i.e.$ a theory for a symmetric rank-2 tensor field $h_{\mu\nu}(x)$ with {\it five dof only}.\\ \  
\\
\textbf{Degrees of freedom :}
In order to determine the dof=5 of any solution, we have to investigate which of the eight solutions satisfy, with or without FP point,  the following two conditions (in momentum space) 
\begin{eqnarray}
p_\nu\tilde{h}^{\mu\nu} &=& 0 \label{4.1} \\
\tilde{h} &=& 0 \label{4.2}
\end{eqnarray}
The first constraint reduces number of parameters(dof) by D in D-dimensions as it is a vector constraint. The second reduces it by $1$ as it is a scalar constraint. Hence, in the  dimension $D=4$ with constraints above for symmetric rank-2 tensor we can obtain dof$=5$ as shown below~\cite{20,21}
\begin{equation}
\left.\underbrace{\frac{D(D+1)}{2}}_{\text{$h_{\mu\nu}=h_{\nu\mu}$}}
-\underbrace{1}_{\text{$h=0$}}
-\underbrace{D}_{\text{$\partial_\mu h^{\mu\nu}=0$}}
=
2s+1\right|_{D=4;s=2}=5.
\label{2.15}\end{equation}

We now proceed to identify solutions which can satisfy both conditions in Eqs.~\eqref{4.1},\eqref{4.2}.  To do so, we need to find equations of motion, which we summarize briefly as given in~\cite{20,21}.
These equations for fields $h_{\mu \nu}$ and $b_\mu$ in the theory given by  Eq.~\eqref{S} are
\begin{eqnarray}
\frac{\delta S'}{\delta\tilde{h}_{\mu\nu}}
&=&
\frac{\delta S_{inv}}{\delta\tilde{h}_{\mu\nu}}+\frac{\delta S_{gf}}{\delta\tilde{h}_{\mu\nu}} + \frac{\delta S_{m}}{\delta\tilde{h}_{\mu\nu}} \nonumber \\
&=&
[(p^2-m_2^2)\eta^{\mu\nu}-p^\mu p^\nu]\tilde{h} - 
(p^2+m_1^2)\tilde{h}_{\mu\nu} \nn\\ 
&-&\frac{i}{2}(p^\mu\tilde{b}^\nu+p^\nu\tilde{b}^\mu)
-i\kappa_1\eta^{\mu\nu}p_\lambda\tilde{b}^\lambda
=0 \label{4.3}\\
\frac{\delta S'}{\delta\tilde{b}_{\mu}} 
&=&
\kappa\tilde{b}^\mu + ip_\nu\tilde{h}^{\mu\nu} + i\kappa_1p^\mu\tilde{h} = 0, \label{4.4}
\end{eqnarray}
 and $\tilde{b}_{\mu}(x)$ is the Fourier transform of ${b}_{\mu}(x)$.

Contracting Eq.~\eqref{4.3} with $\eta_{\mu\nu}$ and $e_{\mu\nu}\equiv \frac{p_\mu p_\nu}{p^2}$,  we get
\begin{eqnarray}
\eta_{\mu\nu}\frac{\delta (S_{inv}+S_{m})}{\delta\tilde{h}_{\mu\nu}} 
&=& 
i(1+D\kappa_1)p_\lambda\tilde{b}^\lambda \label{4.5} \\
e_{\mu\nu}\frac{\delta (S_{inv}+S_{m})}{\delta\tilde{h}_{\mu\nu}}
&=&  
-i(1+\kappa_1)p_\lambda\tilde{b}^\lambda \label{4.6}.
\end{eqnarray}
The idea is to check whether any of the eight solutions in Eqs.~(\ref{3.4}-\ref{3.11}), when put in Eqs.~(\ref{4.5}, \ref{4.6})   leads to the constraints in Eqs.~(\ref{4.1},\ref{4.2}) needed to have the desired five dof of massive gravity.

Solution 2 in Eq.~\eqref{3.5}  satisfies the constraint in Eqs.~(\ref{4.1}, \ref{4.2}) and hence it is the massive gravity solution at FP point $m_1^2+m_2^2=0$~\cite{20,21}. Let us  illustrate this for sake of self consistency of the work. Putting the values for the parameters $m_1^2$, $m_2^2$, $\kappa$ and $\kappa_1$  of solution 2, Eqs. (\ref{4.4}, \ref{4.5} and \ref{4.6}) 
become, respectively 
\begin{equation}
\kappa\tilde{b}^\mu + ip_\nu\tilde{h}^{\mu\nu} - ip^\mu\tilde{h} = 0 
\label{4.7}\end{equation}
\begin{equation}
[p^2(D-2)-(m_1^2+Dm_2^2)]\tilde{h} + (2-D) p^2e_{\mu\nu}\tilde{h}^{\mu\nu} 
= i(1-D)p_\mu\tilde{b}^\mu 
\label{4.8} \end{equation}
\begin{equation}
m_2^2\tilde{h}+m_1^2e_{\mu\nu}\tilde{h}^{\mu\nu} = 0 
\label {4.9}\end{equation}
As $m_2^2\neq 0$, from Eq.~\eqref{4.9} we get
\begin{equation}
\tilde{h}=-\frac{m_1^2}{m_2^2}e_{\mu\nu}\tilde{h}^{\mu\nu}
\label{4.10}\end{equation}
hence, by Eq.~\eqref{4.7}
\begin{equation}
\tilde{b}^\mu=-\frac{i}{\kappa}\frac{m_1^2+m_2^2}{m_2^2}p_\nu\tilde{h}^{\mu\nu}.
\label{4.11}\end{equation}
Substituting Eqs.~(\ref{4.10},\ref{4.11}) into Eq.~\eqref{4.8}, we have
\begin{equation}
(m_1^2+m_2^2)(D-2+\frac{1-D}{\kappa})p^2e_{\mu\nu}\tilde{h}^{\mu\nu}=(m_1^2+Dm_2^2)e_{\mu\nu}\tilde{h}^{\mu\nu}
\label{4.12}\end{equation}
For the solution 2, the term $D-2+\frac{1-D}{\kappa}\neq0$ since $\kappa\leq\frac{D-1}{2-D}$ and $m_1^2+Dm_2^2<0$. Hence if and only if the FP condition, $m_1^2+m_2^2=0$ is met, from Eqs.~(\ref{4.7}, \ref{4.10}  and \ref{4.11}) we get respectively 
\be
\tilde{h}=0 \nn\\
p_\nu\tilde{h}^{\mu\nu}=0 \label{4.14}.
\ee

These are the necessary conditions to have 5 dof, according to Eq.~\eqref{2.15}.  Elsewhere outside the FP point in the generic range of masses, no easy identification of the dof  can be obtained for combination b of parameters $\kappa$ and $\kappa_1$. Therefore the solutions $3, 4, 5, 6$  are not massive gravity solutions as ranges of masses $m_1^2$ and $m_2^2$ in them do not include FP point. Similarly solutions $1$ and $7$,    with or without FP point, fail to satisfy either both or one of the constraints in Eqs.~(\ref{4.1}, \ref{4.2}).  Hence, they also do not represent massive gravity.

The remaining solution 8 in Eq.~\eqref{3.11} gives $dof=5$ outside the FP point~\cite{20,21}. To check this,  
we write the general equations \eqref{4.4} and \eqref{4.6} for the solution 8 as follows 
\begin{equation}
p_\nu\tilde{h}^{\mu\nu} - p^\mu\tilde{h} = 0 
\label{4.16}\end{equation}
\begin{equation}
m_2^2\tilde{h}+m_1^2e_{\mu\nu}\tilde{h}^{\mu\nu} = 0.
\label {4.17}\end{equation}
From the above two Eqs.  we get
\begin{equation}
(m_1^2+m_2^2)\tilde{h}=0.
\label{4.19}\end{equation}
This condition is satisfied either on the FP point $m_1^2+m_2^2=0$, or else outside the FP point, provided that 
\begin{equation}
\tilde{h}=0,
\label{4.20}\end{equation}
which, substituted into Eq.~\eqref{4.16}, gives 
\begin{equation}
p_\nu\tilde{h}^{\mu\nu}=0.
\label{4.21}\end{equation}
Hence for solution 8, we get the required  dof  outside the FP point. In conclusion, we learnt that there are two massive gravity solutions, solution 2 at FP point and solution 8 outside the FP point.  In the next section we  show that these two seemingly discrete physical solutions valid in two different regimes of masses are actually equivalent at the level of generating functional. 
\section{EQUIVALENCE OF TWO SOLUTIONS Under FFBRST}
   
Now we briefly outline the procedure for the passage from the  BRST  to  FFBRST transformations that will ease to understand the work of paper. We start  with making the 
infinitesimal global parameter $\delta\omega$ field dependent  by introducing a numerical parameter $ \beta  \ (0\leq \beta\leq 1) $ and 
making all the fields  $\beta $ dependent
such that  $\phi (x,\beta=0)=\phi(x) $ and $\phi (x,\beta=1)=\phi^\prime(x) $, the transformed field.
The symbol $\phi$ generically describes all the fields  $b_\mu,\xi_\mu,\bar \xi_\mu$.  The BRST transformation in Eq. \eqref{4} is then written as
\be\label{infb}
d\phi = \delta_b[\phi(x,\beta)]\Theta^\prime(\phi(x,\beta))\ d\beta
\ee
where $\Theta'$ is a finite field dependent anti-commuting parameter and 
$\delta_b[\phi(x,\beta)]$ is the form of the transformation for the corresponding field as in 
Eq.~\eqref{4}. The
FFBRST is then constructed by integrating Eq. \eqref{infb} from $\beta=0$ to $\beta=1$ as 
\be\label{ffbrst}
\phi^\prime\equiv \phi(x,\beta=1)=\phi (x,\beta=0)+\delta_b[\phi(0)]\Theta [\phi(x)]
\ee
where $\Theta[\phi(x)] =\int_0^1 d\beta^\prime\Theta^\prime[\phi(x,\beta)] $. Like usual BRST transformation,  FFBRST
transformation leaves the effective action  invariant. However, since the transformation parameter is field dependent in nature, 
FFBRST transformation does not leave the path integral measure, ${\cal D}\phi $ invariant and produces a non-trivial Jacobian factor $J$. 
This $J$  can further be cast as a local  functional of fields, $ e^{iS_J}$ (where the $S_J$ is the action representing the Jacobian factor $J$) 
if the following condition is met \cite{22}
\be \label{con}
\int { \cal D }\phi (x,\beta) \left [\frac{1}{J}\frac{dJ}{d\beta}-i\frac{dS_J}{d\beta}\right ]e^{i(S_J+\mathcal{S}_{eff})}=0.
\ee

Thus the procedure for FFBRST may be summarised as (i)  calculate the infinitesimal change in 
Jacobian, $\frac{1}{J}
\frac{dJ}{d\beta} d\beta $ using 
\begin{equation}
\frac{J(\beta)}{J(\beta+d\beta) }= 1-\frac{1}{J(\beta)}\frac{dJ(\beta)}{d\beta}d\beta
= \sum_\phi \pm \frac{\delta\phi(x,\beta+d\beta)}{\delta\phi(x,\beta)}
\end{equation}
for infinitesimal BRST transformation, $+$ or $-$ sign is for Bosonic or Fermion nature of the field  $\phi$ respectively 
(ii)  make an ansatz for $S_J$, (iii)  then  prove the Eq. (\ref{con})
for this ansatz and finally (iv) replace $J(\beta)$ by $e^{iS_J}$ in the generating functional
\begin{equation}
W=\int {\cal D}\phi (x) e^{iS_{eff}(\phi)} = \int {\cal D}\phi (x,\beta) J(\beta)e^{iS_{eff}(\phi (x,\beta))} .
\label{ww}
\end{equation}
Setting $\beta=1$, this would then provide the new effective action $ S^\prime_{eff}=S_J+S_{eff}$.\\
Having given the general structure of FFBRST we are now in a position to move on to the main objective of the paper.
In order to connect two theories,  the initial theory has to be BRST invariant.  Here in the present case two theories are that of solutions 2 and  8. We begin with the action of solution 2 which in $D=4$ is given as below
\begin{eqnarray}
S_2&=& S_{inv}+\int d^4x[b^\mu(\partial^\nu h_{\mu\nu} - \partial_\mu h)+\frac{\kappa }{2} b_\mu b^\mu] +\int d^4x \ m_1^2(h_{\mu\nu} h^{\mu\nu}-h^2)+S_{ghost},
\end{eqnarray}
where $\kappa\leq\frac{-3}{2}$ for $D=4$ and $m_1^2 > 0 m_1^2 +D m_2^2 <0$.	
We see that the mass term is written at FP point since this solution is physical massive gravity only at FP point.
The action $S_2$ is BRST invariant under following transformation
\begin{eqnarray}\label{tra}
\hat{s} h_{\mu\nu} &=& (\partial_\mu\xi_\nu + \partial_\nu\xi_\mu)\  \delta \omega\nn\\
\hat{s} \xi_\mu &=& 0  \\
\hat{s} \bar\xi_\mu &=& b_\mu \ \delta \omega\nn\\
\hat{s} b_\mu &=&2m_1^2\xi_\mu \ \delta \omega\nn,
\end{eqnarray}
subjected to the harmless condition
\begin{eqnarray}\label{b}
b^\mu \xi_\mu=0.
\end{eqnarray} 
The $b^\mu$ is an auxiliary field which does not propagate,  fields $\xi_\mu, \overline{\xi_\mu}$ are ghosts and constraint in Eq.~\eqref{b} does not involve any derivative. Hence,  this condition does not alter the dynamics of the theory of solution 2. Therefore, we are safe to impose it, which can be obtained by putting the following constraint on ghosts
\begin{eqnarray}\label{tra2}
\bar \xi^\mu \xi_\mu=0
\end{eqnarray}
Thus, Eq.~\eqref{tra2} will be useful in eliminating one more undesired term as we see shortly.
The BRST variation of Eq.~\eqref{tra2}  under the transformation\eqref{tra} gives us the condition required for BRST invariance of solution 2 i.e.,  
\begin{eqnarray}
\hat{s}(\bar \xi^\mu \xi_\mu)=b^\mu \xi_\mu=0.
\end{eqnarray}
Thus, we note that there is only one independent condition of Eq.~\eqref{tra2}.
 Now, the action of the other solution  8 ($\kappa =0 $) is given by
\be \label{har}
S_8=S_{inv}+\int d^4x\ b^\mu(\partial^\nu h_\mu\nu - \partial_\mu h)+\int d^4x\ [m_1^2 h^{\mu\nu}  h_{\mu\nu}+ m_2^2 h^2]+ S_{ghost},
\ee
with $ m_1^2  \geq 0, m_1^2 + D m_2^2 \neq 0 $. Because this solution is the physical massive gravity outside the FP point, we have put the general mass term in the action. 
To establish the equivalence of the two solutions, we resort to FFBRST and demonstrate that generating functional corresponding to the solution 2 can be connected to that  of solution 8 through appropriately constructed FFBRST transformation.
For the proposed connection, we consider the following action
\be \label{ini}
S&=&S_{inv}+\int d^4x[b^\mu(\partial^\nu h_{\mu\nu} - \partial_\mu h)+\frac{\kappa }{2} b_\mu b^\mu] +\int d^4x \ m_1^2(h_{\mu\nu} h^{\mu\nu} - h^2)\nn \\&+&\int d^4 x\ (m_1^2+m_2^2)h^2 + S_{ghost}.
\ee
At FP point, second last vanishes and hence we see that  $S=S_2$.
 
Now we make all the fields of action $S$ numerical parameter $\beta$ dependent, consider $S$ at FP point and  then apply FFBRST transformation $\hat{s}_{FF}$ of the form as in Eq.~\eqref{tra} with appropriate 
finite field dependent parameter in Eq.~\eqref{thita} on the corresponding generating functional $W_2 = \int \mathcal{D}\phi\  e^{iS_{at FP}}$. Symbol $\phi$ generically represents all fields in the theory. The action in Eq.~\eqref{ini} at FP point remains invariant but the path integral measure changes and  will produce an additional local contribution as in Eq.~\eqref{sj}, which will add in Eq.~\eqref{ini} to give the action representing solution 8, Eq.~\eqref{har} when we move away from FP point. Symbolically we can represent the stated process as follows
\be\label{49}
\hat{s}_{FF} W_2 \rightarrow \int \mathcal{D} \phi\  e^{S_{at FP}+ S_J}  \xrightarrow{\text{away from FP point,} m_1^2+m_2^2 \neq 0} W_8\   \  (\text{at} \ \beta =1 )
\ee
 Thus we would achieve our goal of showing equivalence between solutions 2 and 8.  We now construct the suitable FFBRST parameter  $\Theta ^{'}$ which will perform the required task mentioned above as follows
\begin{eqnarray}\label{thita}
\Theta^{'}[\phi(x, \beta)] =i\int d^{4}x\   \gamma \bar \xi ^\mu(x, \beta) b_\mu(x, \beta),
\end{eqnarray}
where $\gamma$ is the arbitrary constant.
This form of $\Theta'$ is of course familiar. It is the tensorial extension of the FFBRST shown in Ref.~\cite{22}, which is required to change the gauge parameter by the finite amount, ($\lambda \rightarrow \lambda'$) in the Lorenz gauge.  Thus, the present work manifests the applicability of the FFBRST discussed in Ref.~\cite{22} to a real problem namely, connecting two physical solutions of massive gravity. 
The corresponding change in Jacobian  can be calculated as
\begin{eqnarray}
\frac{1}{J}\frac{dJ}{d\beta}&=&\int d^{4}y\ [(s_bb_\mu)\frac{\delta  \Theta^{'}}{\delta b_\mu}-(s_b \bar \xi_\mu)\frac{\delta \Theta^{'}}{\delta \bar \xi_\mu}]\nn\\
&=&i\int d^4y\ \gamma(2{m_1^2} \xi_{\mu} \bar \xi^{\mu} - b_{\mu} b^{\mu})
\shortintertext{which because of  constraint in Eq.~\eqref{tra2} becomes }
&=&-i\int d^4{y} \ \gamma b^{\mu}b_{\mu}
\end{eqnarray}
Therefore, obvious ansatz for $S_J$ is,
\begin{eqnarray}\label{sj}
S_J&=&\int d^{4}y \ \alpha(\beta)  b^{\mu}b_{\mu},
\end{eqnarray}
where $\alpha(\beta)$ is an arbitrary function
with the initial condition
\be
\alpha(\beta=0)=0.
\ee
 This is to make sure that $S_J=0$ when the transformation has not been applied.
It is easy to see that the condition for the existence of $S_J$ in Eq.~(\ref{con}) is obeyed only if
\begin{eqnarray}\label{ddd}
\frac{d \alpha}{d \beta}=-\gamma \implies \alpha= -\gamma\beta
\end{eqnarray}
We choose $\gamma=\kappa/2$,
where $\kappa \leq \frac{-3}{2}$ for $D=4$,  so that as suggested by Eq.~\eqref{49} adding $S_J$ to the action in Eq.~\eqref{ini} at FP point, we get at $\beta=1$
\begin{eqnarray}
S + S_J &=& S_{inv}+\int d^4x[b^\mu(\partial^\nu h_\mu\nu - \partial_\mu h)+\frac{\kappa }{2} b_\mu b^\mu] +\int d^4x m_1^2(h_{\mu\nu} h^{\mu\nu}-h^2)\nn\\ &-&  \int d^4x \frac{\kappa}{2} b^{\mu}b_{\mu}
\shortintertext {Now as we drift away from the FP point the second last term in Eq.~\eqref{ini} fires and therefore we have }&=&
S_{inv}+\int d^4x[b^\mu(\partial^\nu h_{\mu\nu} - \partial_\mu h) +\int d^4x \ m_1^2(h_{\mu\nu} h^{\mu\nu} - h^2)\nn +\int d^4 x\ (m_1^2+m_2^2)h^2 + S_{ghost}\nn\\
&=& S_{inv}+\int d^4x \ b^\mu(\partial^\nu h_\mu\nu - \partial_\mu h)+\int d^4x\ (m_1^2 h^{\mu\nu}  h_{\mu\nu}+ m_2^2 h^2)+ S_{ghost}\nn 
\shortintertext{which represents the solution 8, a physical massive gravity outside FP point} &=& S_8
\end{eqnarray}     
Thus, we achieve our aim of the paper.

\section{Conclusion}
In this paper we considered linearized gravity as an ordinary gauge field theory, written in terms of rank-2 symmetric tensor $ h_{\mu\nu}(x)$, with gauge fixing and the most general mass term.  We saw that there are only two massive gravity solutions as they only satisfy requirements of being a physical theory mentioned at the beginning of section III. We found that one of the two solution $S_2$ is physical gravity at FP point and other $S_8$   is physical gravity outside the FP point. Thus,  at first they seem unrelated as they both belong to different regimes of masses.  The significance of this work is that we have been able to show,  with the certain constraint on ghosts (Eq.~\eqref{tra2}), the  equivalence at the level of generating functional between these two solutions under FFBRST. The present work provides the practical application of the FFBRST shown in Ref.~\cite{22} to a physical problem. 

\acknowledgements
Haresh sincerely acknowledges the support of SERB, Department of Science and Tech., India for National Postdoctoral Fellowship program with grant no. `PDF/2017/000066'.

\end{document}